\documentclass[aps,superscriptaddress,superbib,twocolumn,groupedaddress]{revtex4}

\usepackage[latin1]{inputenc}
\usepackage[T1]{fontenc}

\RequirePackage{ifthen}%
\RequirePackage[cdot,textstyle,amssymb]{SIunits}%
\RequirePackage{units}
\RequirePackage{color} \RequirePackage{xspace} \RequirePackage{upgreek} \RequirePackage{amsmath} \RequirePackage{amssymb} 
\RequirePackage{wasysym}
\RequirePackage[mathcal]{euscript}
\RequirePackage{eurosym}
\RequirePackage{textcomp}

\RequirePackage{ulem} 

\RequirePackage{graphicx}
\RequirePackage{wrapfig}

\RequirePackage{hyperref}
\definecolor{link_green}{rgb}{0.0,0.7,0.0}
\definecolor{link_blue_dark}{rgb}{0.0,0.0,0.7}
\definecolor{link_blue}{rgb}{0.0,0.0,1}
\definecolor{link_red}{rgb}{0.7,0.0,0.0}
\definecolor{link_red_dark}{rgb}{0.6,0.0,0.0}
\definecolor{link_red_verydark}{rgb}{0.3,0.0,0.0}
\hypersetup{hyperfigures=true,hyperfootnotes=true,hyperindex=true,pagebackref,breaklinks=true,citecolor=link_green,filecolor=link_blue,linkcolor=link_blue_dark, 
urlcolor=link_blue,colorlinks,bookmarks,bookmarksopen,bookmarksnumbered, pdfauthor="Olivier Fruchart",pdfstartview=FitH} 
\RequirePackage[all]{hypcap}

    \newcommand{\remove}[1]{}
    \newcommand{\softremove}[1]{}

    \newcommand{\incharge}[1]{}%

  \providecommand\OFcomment[1]{} 
  \newcommand\finalize[1]{}
  \newcommand\improvement[1]{} 
  \newcommand\addFIG[1]{}
  \newcommand\livelecture[1]{}
  \newcommand\futurecite[1]{}%
  \newcommand\futureref[1]{}%
  \newcommand\futurefig[1]{}%
  \newcommand\futuretab[1]{}%
  \newcommand\dataref[1]{}%

  \definecolor{olivier_comment}{rgb}{0.6,0.6,0.9}
  \newcommand{\olivier}[1]{}
  \newcommand{\olivierComment}[1]{{}}
  \newcommand{\olivierRemove}[1]{{}}
  \newcommand{\olivierReplace}[2]{{}}

\newcommand{\subfigref}[2]{\figurename~\ref{#1}#2}%
\newcommand{\figref}[1]{\subfigref{#1}{}}%
\newcommand{\bracketsubfigref}[2]{~(\figref{#1}#2)}
\newcommand{\bracketfigref}[1]{\bracketsubfigref{#1}{}}

\def\Ms{M_{\mathrm s}}
\def\ExchangeLength{\Delta_\mathrm{d}}
\def\DipolarExchangeLength{\ExchangeLength}
\def\muZero{\mu_{\mathrm 0}}%

\def\muB{\ifmmode{\mu_\mathrm{B}}\else{$\mu_\mathrm{B}$}\fi}%

\def\saphir{\ifmmode{\mathrm{Al}_2\mathrm{O}_3}\else{Al$_2$O$_3$}\fi}%

\def\gammaint{\gamma_\mathrm{int}}

 \newcommand{\vect}[1]{\mathbf{#1}}
\newcommand{\submathrm}[1]{_{\mathrm{#1}}}
\def\micron{\ifmmode{\upmu\mathrm{m}}\else{$\upmu$m}\fi}

\def \AA {\hbox{\r A}}
\newcommand{\lengthnm}[1]{\unit[#1]{nm}}%
\def\deg{\ifmmode{^{\circ}}\else{$^{\circ}$}\fi}%
\def\LLNdeg{\ifmmode{^{\circ}}\else{$^{\circ}$}\fi}%
\def\degC{\LLNdeg\ifmmode{\mathrm C}\else C\fi}%
\newcommand\angledeg[1]{#1\LLNdeg}
\def\kb{\ifmmode{k\submathrm{B}}\else{$k\submathrm{B}$}\fi}
\def\SI{\ifmmode{\mathrm{S.I.}}\else{S.I.}\fi} 
\newcommand{\tempC}[1]{\unit[#1]{\degC}}%

\newcommand{\unitmatrix}[4]{\ifmmode{[#1\;#2\;#3\;#4]}\else{$[#1\;#2\;#3\;#4]$}\fi}

\def\ie{{\it i.e.}\/\xspace}%
\def\etal{{\it et~al.}\/\xspace}%
\def\apriori{\textsl{a~priori}\/\xspace}%
\newcommand\newabbr[3]{
  \expandafter\def\csname#1\endcsname{#2}
  \expandafter\def\csname#1ACRONYM\endcsname{#1}
  \expandafter\def\csname#1SHORT\endcsname{#2}
  \expandafter\def\csname#1LONG\endcsname{#3}
}

\newcommand\newjournal[3]{\newabbr{#1}{#2 }{#3 }}

\newjournal{acie}{Angew. Chem. Int. Ed.}{Angew. Chem. Int. Ed.}%
\newjournal{ac}{Act. Cryst.}{Acta Crystallographica}%
\newjournal{acp}{Adv. Chem. Phys.}{Advances in Chemical Physics}%
\newjournal{ACSNano}{Am. Chem. Soc. Nano}{American Chemical Society Nano}%
\newjournal{ActaMater}{Acta. Mater.}{Acta Materiala}%
\newjournal{AdvMater}{Adv. Mater.}{Advanced Materials}%
\newjournal{AdvPhys}{Adv. Phys.}{Advances in Physics}%
\newjournal{afm}{Adv. Funct. Mater.}{Advanced Functional Materials}%
\newjournal{AIPAdv}{AIP Advances}{AIP Advances}%
\newjournal{AIPcp}{AIP Conf. Proc.}{AIP Conferences Proceedings}%
\newjournal{am}{Adv. Mater.}{Advanced Materials}%
\newjournal{amm}{Acta Metall. Mater.}{Acta Metallica Materia}%
\newjournal{AnnPhys}{Ann. Phys.}{Annales de Physique}%
\newjournal{ap}{Appl. Phys.}{Applied Physics}%
\newjournal{apa}{Appl. Phys. A}{Applied Physics A: Materials Science and Processing}%
\newjournal{apex}{Appl. Phys. Express}{Applied Physics Express}%
\newjournal{apl}{Appl. Phys. Lett.}{Applied Physics Letters}%
\newjournal{ApplPhys}{Appl. Phys.}{Applied Physics}%
\newjournal{apr}{Appl. Phys. Rev.}{Applied Physics Reviews}%
\newjournal{ass}{Appl. Surf. Sci.}{Applied Surface Science}%
\newjournal{bsfp}{Bull. Soc. Fr. Phys.}{Bulletin de la Soci\'{e}t\'{e} Fran\c{c}aise de Physique}%
\newjournal{CahPhys}{Cah. Phys.}{Cahier de Physique}%
\newjournal{ChemMater}{Chem. Mater.}{Chemistry of Materials}%
\newjournal{cap}{Curr. Appl. Phys.}{Current Appleid Physics}%
\newjournal{ChinPhys}{Chin. Phys.}{Chinese Physics}%
\newjournal{cjp}{Czechols. J. Phys.}{Czecholsovak Journal of Physics}%
\newjournal{cms}{Comput. Mater. Sci.}{Computational Materials Science}%
\newjournal{CollSFN}{Collection SFN}{Collection SFN}%
\newjournal{cp}{Contemp. Phys.}{Contemporary Physics}%
\newjournal{cras}{C. R. Acad. Sci.}{Compte-Rendus de l'Acad\'{e}mie des Sciences}%
\newjournal{crp}{C. R. Physique}{Compte-Rendus de Physique}%
\newjournal{crt}{Cryst. Res. Technol.}{Crystal Research and Technology}%
\newjournal{danSSSR}{Dok. Akad. Nauk SSSR}{Dok. Akad. Nauk SSSR}%
\newjournal{ds}{Dat. Stor.}{Data Storage}%
\newjournal{epjap}{Europhys. J.: Appl. Phys.}{Europhysics Journal: Applied Physics}%
\newjournal{epjb}{Europhys. J. D}{Europhysics Journal D}%
\newjournal{epjd}{Europhys. J. D}{Europhysics Journal D}%
\newjournal{epn}{Europhys. News}{Europhysics News}%
\newjournal{essl}{Electrochem. Solid-State Lett.}{Electrochemical and Solid-State Letters}%
\newjournal{el}{Electron. Lett.}{Electronics Letters}%
\newjournal{epl}{Europhys. Lett.}{Europhysics Letters}%
\newjournal{IEEEted}{IEEE Trans. Electr. Dev.}{IEEE Transactions on Electronic Devices}%
\newjournal{IEEEtm}{IEEE Trans. Magn.}{IEEE Transactions on Magnetics}%
\newjournal{IBMjrd}{IBM J. Res. Develop.}{IBM Journal of Research and Development}%
\newjournal{ijmpb}{Int. J. Mod. Phys. B}{International Journal of Modern Physics B}%
\newjournal{inc}{Il Nuov.Cim.}{Il Nuovo Cimento}%
\newjournal{InterfSci}{Interf. Science}{Interface Science}%
\newjournal{intermetallics}{Intermetallics}{Intermetallics}%
\newjournal{izk}{Int. Zeit. Kristallogr.}{International Zeitschrift f\"{u}r Kristallography}
\newjournal{jac}{J. Appl. Cryst.}{Journal of Applied Crystallography}%
\newjournal{jacs}{J. Am. Chem. Soc.}{Journal of the American Chemical Society}%
\newjournal{jap}{J. Appl. Phys.}{Journal of Applied Physics}%
\newjournal{jasa}{J. Acoust. Soc. Am.}{Journal of the Acoustical Society of America}%
\newjournal{jcg}{J. Cryst. Growth}{Journal of Crystal Growth}%
\newjournal{jcp}{J. Comput. Phys.}{Journal of Computational Physics}%
\newjournal{jes}{J. Electrochem. Soc.}{Journal of The Electrochemical Society}%
\newjournal{jemt}{J. Electron Microsc. Tech.}{}%
\newjournal{jesrp}{J. Electron Spectr. Rel. Phenom.}{Journal of Electron Spectroscopy and Related Phenomena}
\newjournal{jetp}{J. Exp. Theor. Phys.}{Journal of experimental and Theoretical Physics}%
\newjournal{jetpl}{J. Exp. Theor. Phys.: Letters}{Journal of experimental and Theoretical Physics: Letters}%
\newjournal{jgr}{J. Geophys. Res.}{Journal of Geophysical Research}%
\newjournal{jjap}{Jap. J. Appl. Phys.}{Japanese Journal of Applied Physics}%
\newjournal{jmmm}{J. Magn. Magn. Mater.}{Journal of Magnetism and Magnetic Materials}%
\newjournal{jmsj}{J. Magn. Soc. Japan}{Journal of the Magnetic Society of Japan}%
\newjournal{jmnmm}{J. Micro./Nanolith. MEMS MOEMS}{Journal of Micro/Nanolithography, MEMS and MOEMS}%
\newjournal{jnr}{J. Nanopart. Res.}{Journal of Nanoparticle Research}%
\newjournal{jpcl}{J. Phys. Chem. Lett.}{Journal of Physical Chemistry Letters}%
\newjournal{jpcm}{J. Phys.: Condens. Matter}{Journal of Physics: Condensed Matter}%
\newjournal{jpcs}{J. Phys.: Conf. Series}{Journal of Physics: Conference Series}%
\newjournal{JourPhys}{J. Phys.}{Journal de Physique}%
\newjournal{jpcb}{J. Phys. Chem. B}{Journal of Physical Chemistry~B}%
\newjournal{jpcc}{J. Phys. Chem. C}{Journal of Physical Chemistry~C}%
\newjournal{jpcs}{J. Phys. Chem. Sol.}{Journal of Physics and Chemistry of Solids}%
\newjournal{jpd}{J. Phys. D: Appl. Phys.}{Journal of Physics D: Applied Physics}%
\newjournal{jpdap}{J. Phys. D: Appl. Phys.}{Journal of Physics D: Applied Physics}%
\newjournal{jpf}{J. Phys. F: Metal Phys.}{Journal of Physics D: Metal Physics}%
\newjournal{jpI}{J. Phys. I}{Journal of Physics I}%
\newjournal{jpIV}{J. Phys. IV: proc.}{Journal of Physics IV: proceedings}%
\newjournal{jpr}{J. Phys. Rad.}{Journal de Physique du Radium}%
\newjournal{jpsj}{J. Phys. Soc. Jap.}{Journal of the Physical Society of Japan}%
\newjournal{jsp}{J. Comput. Phys.}{Journal of Computational Physics}%
\newjournal{jsgst}{J. Sol-Gel Sci. technol.}{Journal of Sol-Gel Science and Technology}%
\newjournal{jvst}{J. Vac. Sci. Technol.}{Journal of Vacuum and Science Technology}%
\newjournal{jvsta}{J. Vac. Sci. Technol. A}{Journal of Vacuum and Science Technology A}%
\newjournal{jvstb}{J. Vac. Sci. Technol. B}{Journal of Vacuum and Science Technology B}%
\newjournal{MaterToday}{Mater. Today}{Materials Today}%
\newjournal{me}{Microelec. Engineer.}{Microelectronics Engineering}%
\newjournal{mrssp}{Mat. Res. Soc. Symp. Proc.}{Material Research Society Symposium Proceedings}%
\newjournal{mseb}{Mater. Sci. Engrg. B}{Materials Science and Engineering B}%
\newjournal{nanolett}{Nano Lett.}{Nano Letters}%
\newjournal{NanoRes}{Nano Res.}{Nano Research}%
\newjournal{Nanoscale}{Nanoscale}{Nanoscale}%
\newjournal{nanotech}{Nanotechnology}{Nanotechnology}%
\newjournal{Nanotech}{Nanotechnology}{Nanotechnology}%
\newjournal{nature}{Nature}{Nature}%
\newjournal{NatMater}{Nat. Mater.}{Nature Materials}%
\newjournal{NatNanotech}{Nat. Naotech.}{Nature Nanotechnology}%
\newjournal{NatPhys}{Nat. Phys.}{Nature Physics}%
\newjournal{nimprb}{Nucl. Instr. Meth. Phys. Res. A}{Nuclear Instruments and Methods in Physical Research A}%
\newjournal{nimprb}{Nucl. Instr. Meth. Phys. Res. B}{Nuclear Instruments and Methods in Physical Research B}%
\newjournal{njp}{New J. Phys.}{New Journal of Physics}%
\newjournal{pa}{Physica A}{Physica A}%
\newjournal{pb}{Physica B}{Physica B}%
\newjournal{PhysicaB}{Physica B}{Physica B}%
\newjournal{pccp}{Phys. Chem. Chem. Phys.}{}%
\newjournal{pe}{Physica E}{Physica E}%
\newjournal{pla}{Phys. Lett. A}{Physics Letters A}%
\newjournal{plpls}{Proc. Leeds  Phil. Liter. Soc.}{Proceedings of the Leeds Philosophical and Literature Society}%
\newjournal{PhilMag}{Philos. Mag.}{Philosophical Magazine}%
\newjournal{PhilMagB}{Philos. Mag. B}{Philosophical Magazine B}%
\newjournal{pmm}{Phys. Met. Metall.}{}%
\newjournal{PhysRev}{Phys. Rev.}{Physical Review}%
\newjournal{PhysRep}{Phys. Rep.}{Physics Reports}%
\newjournal{prb}{Phys. Rev. B}{Physical Review B}%
\newjournal{prl}{Phys. Rev. Lett.}{Physical Review Letters}%
\newjournal{PhysSolState}{Phys. Sol. State}{Physics of the Solid State}%
\newjournal{ptrsla}{Phil. Trans. Roy. Soc. Lond. A}{Philosophical  Transactions of the Royal Society of London A}%
\newjournal{prsla}{Proc. Roy. Soc. Lond. A}{Proceedings of the Royal Society of London A}%
\newjournal{pss}{Phys. Stat. Sol.}{Physica Status Solidi}%
\newjournal{pssa}{Phys. Stat. Sol. (a)}{Physica Status Solidi (a)}%
\newjournal{pssb}{Phys. Stat. Sol. (b)}{Physica Status Solidi (b)}%
\newjournal{pzs}{Phys. Z. Sowjetunion}{Phys. Z. Sowjetunion}%
\newjournal{pms}{ Prog. Mater Sci.}{Progress in Materials Science}%
\newjournal{progss}{Prog. Surf. Sci.}{Progress of Surface Science}%
\newjournal{rmp}{Rev. Mod. Phys.}{Review of Modern Physics}%
\newjournal{rpp}{Rep. Prog. Phys.}{Report of Progress in Physics}%
\newjournal{rsi}{Rev. Sci. Instr.}{Review of Scientific Instruments}%
\newjournal{sia}{Surf. Interf. Analysis}{Surface and Interface Analysis}%
\newjournal{science}{Science}{Science}%
\newjournal{SciRep}{Sci. Rep.}{Scientific Reports}%
\newjournal{sm}{Scripta Mater.}{Scripta Materialia}%
\newjournal{small-journal}{Small}{Small}%
\newjournal{srn}{Synchrotron Radiat. News}{Synchrotron Radiation News}%
\newjournal{ssc}{Sol. State Comm.}{Solide State Communications}%
\newjournal{sss}{Sol. State Sci.}{Solide State Sciences}%
\newjournal{sst}{Supercond. Sci. Technol.}{Superconducting Science and Technology}%
\newjournal{sumi}{Superlattices Microstruct.}{Superlattices and Microstructures}%
\newjournal{srl}{Surf. Rev. Lett.}{Surface Review and Letters}%
\newjournal{surfsci}{Surf. Sci.}{Surface Science}%
\newjournal{ssl}{Surf. Sci. Lett.}{Surface Science Letters}%
\newjournal{ssr}{Surf. Sci. Rep.}{Surface Science Reports}%
\newjournal{tfs}{Trans. Faraday Soc.}{Transactions of the Faraday Society}%
\newjournal{tpl}{Tech. Phys. Lett.}{Technical Physics Letters}%
\newjournal{tsf}{Thin Solid Films}{Thin Solid Films}%
\newjournal{ultramicroscopy}{Ultramicr.}{Ultramicroscopy}%
\newjournal{zm}{Z. Metallkd.}{Zeitschrift f\"{u}r Metallkunde}%
\newjournal{zp}{Z. Phys.}{Zeitschrift f\"{u}r Physik}%
\newjournal{zpc}{Z. Phys. Chem.}{Zeitschrift f\"{u}r Physikalische Chemistry}%

\begin{document}

\renewcommand\topfraction{0.8}
\renewcommand\bottomfraction{0.7}
\renewcommand\floatpagefraction{0.7}


\title{Growth and micromagnetism of self-assembled epitaxial fcc(111) cobalt dots}


\author{O. Fruchart\footnote{Olivier.Fruchart@grenoble.cnrs.fr}}%
\affiliation{Institut N\'{E}EL, CNRS \& Universit\'{e} Joseph Fourier -- BP166 -- F-38042 Grenoble Cedex 9 -- France}%
\author{A. Masseboeuf\footnote{Present address: CEMES, 29 rue Jeanne Marvig, 31055 Toulouse, France}}%
\affiliation{CEA-Grenoble, INAC/SP2M/LEMMA, 17 rue des Martyrs, Grenoble, France}%
\author{J. C. Toussaint}%
\affiliation{Institut N\'{E}EL, CNRS \& Universit\'{e} Joseph Fourier -- BP166 -- F-38042 Grenoble Cedex 9 -- France}%
\affiliation{Grenoble - Institut National Polytechnique -- France}%
\author{Pascale Bayle-Guillemaud}
\affiliation{CEA-Grenoble, INAC/SP2M/LEMMA, 17 rue des Martyrs, Grenoble, France}%

\date{\today}

\begin{abstract}

We developed the self-assembly of epitaxial submicrometer-sized face-centered-cubic~(fcc) Co(111) dots using pulsed laser 
deposition. The dots display atomically-flat facets, from which the ratios of surface and interface energies for fcc Co are 
deduced. Zero-field magnetic structures are investigated with magnetic force and lorentz microscopies, revealing vortex-based 
flux-closure patterns. A good agreement is found with micromagnetic simulations.

\end{abstract}



\maketitle

\vskip 0.5in


\vskip 0.5in


\section{Introduction}

Physical and chemical properties are often significantly changed in low-dimensional systems, compared with materials in their bulk 
form. Top-down techniques based on the combination of lithography and thin-film deposition are the major approach for producing 
low-dimensional structures, because of the nearly unlimited freedom for designing planer shapes. Top-down techniques however reach 
their limits when it comes to producing complex three-dimensional systems, or features with very small sizes. For such purposes 
bottom-up synthesis routes, also called self-assembly, are more effective and their use has been constantly rising in the past one 
or two decades. Dramatic achievements have been made in both directions, with for instance porous templates such as the case of 
anodic alumina\cite{bib-MAS1995,bib-LEE2006,bib-LEE2010b,bib-KNE2007} or programmable architecture based on matching DNA 
strands\cite{bib-ROT2006b} for complex non-flat architectures, and the fabrication of supported clusters with a size controlled 
down to one single atom\cite{bib-GAM2002}.

In the field of magnetism, the use of self-assembly for its ability to produce tiny structures has brought many 
breakthroughs\cite{bib-BAN2005,bib-FRU2005,bib-END2010} such as: giant orbital moment and magnetic anisotropy energy at surfaces of 
itinerant ferromagnets\cite{bib-GAM2002}, ferromagnetism of small Rh clusters while its bulk form is 
non-magnetic\cite{bib-COX1993}; manipulation through electric field of the magnetic order such as antiferromagnetic 
Fe\cite{bib-GER2011}; nanometers-sized ferromagnetic columns embedded in a non-magnetic matrix\cite{bib-JAM2006,bib-VID2009}, 
synthesis of artificial multiferroic metamaterials\cite{bib-ZHE2004}. Concerning thick or even three-dimensional structures, 
self-assembly has been used to produce assemblies of nanowires in porous membranes\cite{bib-FER1999a,bib-NIE2000}, clusters and 
elongated objects through chemistry\cite{bib-CHA2005}, and dots and wires at surfaces by physical means based on dewetting 
processes\cite{bib-FRU2001b,bib-FRU2005b,bib-ZDY2006,bib-ROU2006,bib-FRU2007}. The typical dimensions of such objects range from a 
few nanometers to a few hundreds of nanometers. At this scale one physical issue is micromagnetism (magnetization reversal and the 
arrangement of domains), self-assembly providing a model system.

We reported a series of micromagnetic studies based on self-assembled sub-micrometer-sized epitaxial Fe(110) dots deposited on 
W(110) and Mo(110) surfaces\cite{bib-FRU2007}. These dots are elongated owing to the (110) symmetry, and may thus display a domain 
wall along its length\cite{bib-VAN1984,bib-FRU2003c}. These dots were thus used to observe\cite{bib-FRU2005} and 
manipulate\cite{bib-FRU2009b,bib-FRU2010a} the internal degrees of freedom of Bloch domain walls. The physics of magnetic vortices, 
the one-dimensional analogous of the two-dimensional domain walls, is also prone to a rich physics\cite{bib-SHI2000b,bib-VAN2006}. 
This calls for the availability of self-assembled dots with a high rotational symmetry, suited for the occurrence of magnetic 
vortices.

Here we report the self-assembly of face-centered cubic (fcc) Co dots along the (111) orientation, with a lateral size of the order 
of a few hundreds of nanometers. Prior reports of self-assembled Co dots exist however with a rather flat shape and with a much 
smaller size so that facets cannot be identified and/or vortices do not form\cite{bib-CHE2004,bib-DIN2005,bib-SIL2005b}. Larger 
dots were reported, however with no thorough magnetic characterization\cite{bib-BEN2010}. We first detail the growth procedure, 
then analyze the topography and facets of the dots, and finally examine their zero-field magnetization configuration, revealing the 
occurrence of a single magnetic vortex. Prior to this report, we used these dots to study the dimensionality cross-over between a 
magnetic domain wall and a magnetic vortex\cite{bib-FRU2010c}.

Epitaxial growth was conducted using pulsed-laser deposition in a set of ultra-high vacuum chambers described 
elsewhere\cite{bib-FRU2007}. The laser is a doubled Nd-YAG ($\lambda=\lengthnm{532}$) with a pulse length of the order of 
$\unit[25]{ns}$. Atomic force microscopy~(AFM) and Magnetic force microscopy~(MFM) were performed with a NT-MDT Ntegra microscope 
operated under atmospheric pressure, using either Asylum MFM probes based on AC240TS Olympus cantilevers, or ultrasharp 
SSS-PPP-MFMR probes from Nanosensors. The MFM signal consists of the phase, with the convention of dark (resp. bright) contrast 
indicative of attractive (resp. repulsive) forces. Lorentz Microscopy was performed on a FEI Titan 80-300 fitted with a Lorentz 
lens and a Gatan Imaging filter. The sample was prepared using mechanical polishing and ion milling. Magnetic information is gained 
in out-of-focus conditions, where bright and dark Fresnel contrast arise from the overlapping (resp. spreading) of electrons 
experiencing different Lorentz forces in neighboring domains. Fresnel contrast thus highlights strong magnetization gradients, such 
as magnetic domain walls and vortices. Micromagnetic simulations were performed using feellgood, a home-built code based on the 
temporal integration of the Landau-Lifshitz-Gilbert equation in a finite element scheme, \ie using tetrahedra to discretize the 
dots\cite{bib-ALO2012}. Only exchange and magnetostatic interactions were taken into account, to deal with the present case of 
magnetically-soft dots. The parameters for bulk cobalt were used: $A=\unit[3\times10^{-11}]{\joule\per\meter}$ and 
$\muZero\Ms=\unit[1.7844]{\tesla}$. The tetrahedron size was \unit[4]{\nano\meter} on the average, smaller than the dipolar 
exchange length of Co $\DipolarExchangeLength=\lengthnm{5}$.

\section{Growth and structure}

\dataref{Samples of interest: AR66 (first dots, TP ENSPG); FRU167, TP; ZKB43}

\begin{figure}
  \begin{center}
  \includegraphics[width=84.797mm]{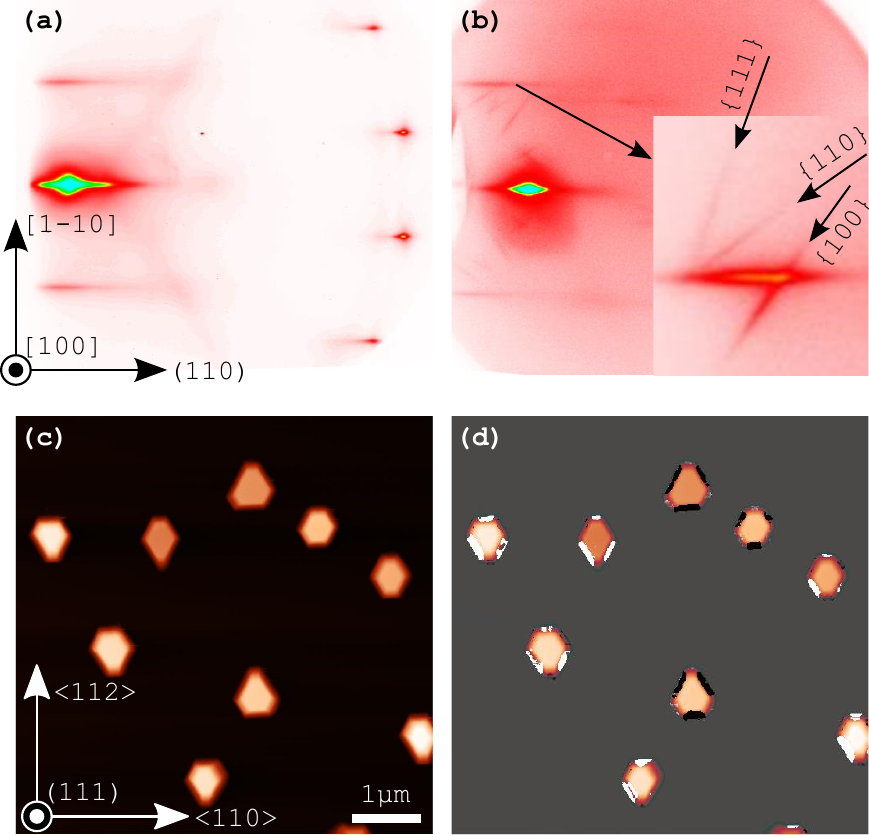}%
  \caption{\label{fig-structure}(a)~RHEED pattern of the W(110) supporting surface with beam azimuth $[100]$\dataref{FRU167, 
W-recuit 022deg q=[001]} (b)~RHEED on Co dots along the same azimuth. The inset is a zoom around the first-order streak, with a 
change of azimuth of a few degrees to better highlight the tilted streaks related to the facets\dataref{AR66}. The angle of the 
arrows is not a fit to the streaks, but is set \apriori to that for fcc planes (c)~AFM image of Co dots\dataref{AR66, TP ENSPG 
2008}. (d)~Same as before, with the ${100}$ facets highlighted in two sets depending on their azimuth: $\angledeg{0}$, 
$\angledeg{120}$ and $\angledeg{240}$~(black); $\angledeg{60}$, $\angledeg{180}$ and $\angledeg{300}$~(white).}
  \end{center}
\end{figure}

We start from a Sapphire $(11\overline{2}0)$ wafer outgassed under UHV. All materials are then evaporated at a rate close to 
$\unit[1]{\AA/min}$. We first deposit $\unit[0.7]{\nano\meter}$ of Mo followed by $\lengthnm{10}$ of W at $\tempC{150}$. A smooth 
and single-crystalline W(110) surface results from annealing this stack at $\unit[900]{\degC}$\cite{bib-FRU2007}. Reflection high 
energy electron diffraction~(RHEED) shows patterns with very narrow peaks perpendicular to the sample surface, indicating the good 
crystalline quality of this buffer layer\bracketsubfigref{fig-structure}a. Scanning tunneling microscopy confirms the flatness of 
the layer, displaying terraces of width several hundreds of nanometers.

Co is then deposited on this surface at $\unit[400]{\degC}$, with a nominal thickness of several nanometers. RHEED patterns are 
invariant upon rotation of the sample by $\angledeg{60}$, which hints at the occurrence of a growth direction either hexagonal 
$(0001)$, or fcc $(111)$. These directions of growth are expected as associated with the densest planes of Co, with a triangular 
symmetry nearly matching in symmetry and dimensions those of body-centered cubic W(110)\cite{bib-BAU1986}. The analysis of the 
RHEED inter-streak distances is consistent with such triangular planes for bulk Co. The absence of $\approx\angledeg{\pm2.5}$ twins 
is indicative of the Nishiyama-Wasserman epitaxial relationship\cite{bib-GOT1989,bib-VAN1994}, for which two sides of the hexagons 
in the Co planes are parallel to the short side of the rectangular surface lattice of~W.

The most stable cristalline structure of bulk Co is hcp below $\tempC{425}$, and fcc above this temperature. Due to an elevated 
temperature and stress related to low dimensions, fcc Co is often found in nanostructures and thin films. The hcp, fcc and even 
bcc\cite{bib-OHT2011} lattices may be selected  depending on the chemical nature and orientation of the supporting surface. Let us 
examine in further detail the RHEED patterns of the deposit to determine its cristalline structure in our case. While patterns 
along the azimuth $\mathrm{W[110]}$ (modulo $\angledeg{60}$) display only streaks perpendicular to the sample surface, patterns for 
the beam azimuth along the $\mathrm{W[001]}$ direction display those streaks, plus a set of streaks tilted at several well-defined 
angles\bracketsubfigref{fig-structure}b. This hints at the occurrence of dots of large size, with a flat top and atomically-flat 
tilted facets\cite{bib-FRU2007}. The set of angles deduced from these patterns is consistent with those expected for 
low-Miller-index planes of a fcc crystal~(see Annex for details, including \figref{fig-facets-annex}). To the contrary 
(calculations not provided here), these angles canot be reproduced based on an hcp structure. This strongly suggests that Co grows 
with a fcc lattice. The epitaxial relationship is thus written: $\mathrm{Co}(111)//\mathrm{W}(110)$. While the occurrence of the 
fcc phase is not surprising due first to the low dimension (usually favoring fcc Co), second to the fact that the stable form of Co 
at the growth temperature is fcc. Nevertheless, we will see in the magnetics part that in practice a very small fraction of dots 
also grows in the hcp structure. These facts are consistent with a report of the growth of Co at elevated temperature on 
$\mathrm{Mg0}(001)$, also yielding self-assembled dots mostly with an fcc structure, in that case with a $(001)$ top facet, and a 
minority with an hcp structure\cite{bib-BEN2010}.

\begin{figure}
  \begin{center}
  \includegraphics[width=88mm]{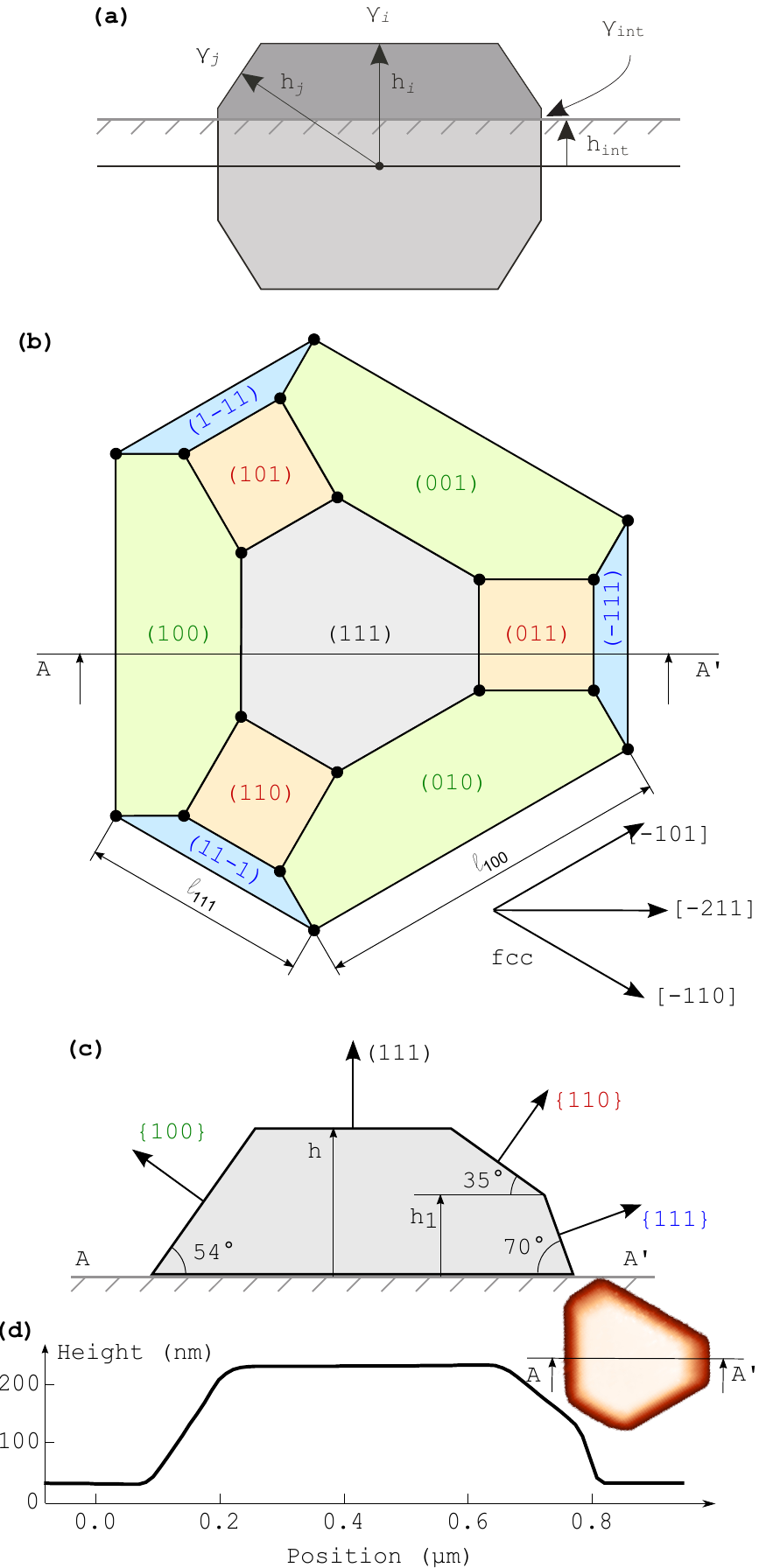}%
  \caption{\label{fig-facets}(a)~Wulff-Kaishev construction, and associated notations (b)~Top-view schematic of a facetted fcc 
$\mathrm{Co}(111)$ dot. The geometrical ratios are chosen based on the mean values of surface and interface energies determined 
experimentally (see text). The in-plane lattice directions are shown in the bottom right corner.  (c)~Schematic and 
(d)~experimental cross-sections. In the latter case the dot used is the one most up in \subfigref{fig-structure}c.}
  \end{center}
\end{figure}

Characterization in real space using AFM in shown in \subfigref{fig-structure}c. The density of the dots is of the order of 
$\unit[0.25]{\micro\reciprocal\meter}$. Like for the case of $\mathrm{Fe}(110)$ dots, this density is mostly determined by 
temperature during deposition, while the thickness of the deposit affects primarily the volume of the dots. AFM confirms the 
smoothness of both the top surface and side facets. Figures for the tilt of the latter are consistent with those deduced from 
RHEED\bracketfigref{fig-facets}. The facet analysis \bracketsubfigref{fig-structure}d shows that the invariance upon rotation of 
$\angledeg{60}$ results from the coexistence of two sets of dots, rotated with each other by $\angledeg{180}$. This twinned 
epitaxial relationship is expected and observed for the fcc(111)/bcc(110) Nishiyama-Wasserman epitaxy. \figref{fig-facets} shows a 
sketch of a typical Co(111) dot resulting from RHEED and AFM data.

The shape of crystals at surfaces is determined by the Wulff-Kaishev construction\cite{bib-HEN2005}, stating that:

\begin{equation}
  \label{eq-WulffKaichev}\frac{\gamma_{i}}{h_{i}} = \frac{(\gamma_\mathrm{S}-\gammaint)}{h_\mathrm{int}} =
  \mathrm{Constant}
\end{equation}

\noindent $h_i$ is the distance from the (possibly hypothetical) center of the crystal to a given set of crystallographic planes of 
the material (the facets), $\gamma_i$ is the free energy of facet~$i$, $\gamma_\mathrm{S}$ is that of the free surface of the 
substrate, $\gammaint$ is the interfacial free energy\cite{bib-HEN2005} (defined as zero for a material in contact with itself), 
and $h_\mathrm{int}$ is a distance related to the vertical aspect ratio of the supported crystal\bracketsubfigref{fig-facets}a. 
Thus ratios of surface and interface energies may be determined from the observed shape of a crystal. We consider as unknowns 
$\Delta\gamma=\gamma_\mathrm{S}-\gamma_\mathrm{int}$, and the energy of the facets identified with RHEED and AFM: $\gamma_{111}$, 
$\gamma_{110}$ and $\gamma_{100}$. Notice that another choice instead of $\Delta\gamma$ would be to use the adhesion energy 
$\gamma^*=\gamma_{111}+\gamma_\mathrm{S}-\gamma_\mathrm{int}$\cite{bib-SIL2005b}.

Only ratios may be determined, which leaves three dimensionless unknowns. Three experimental geometrical ratios are required to 
determine these three figures, see appendix~II. Based on the analysis of all dots on \subfigref{fig-structure}a we find 
$\Delta\gamma=(0.41\pm0.11)\gamma_{111}$, $\gamma_{110}=(1.06\pm0.04)\gamma_{111}$ and $\gamma_{100}=(0.88\pm0.07)\gamma_{111}$. 
The error figures stand for the standard deviation. The larger energy for $\{111\}$ and $\{110\}$ surfaces than for $\{100\}$ 
surfaces is responsible for the triangular shape of our nanocrystals, with the former facets of smaller extent than the latter. 
Notice also the larger standard deviation for the value of $\Delta\gamma$. This is consistent with the sensitivity of adhesion 
energy on adsorbates such as resulting from residual gas\cite{bib-FRU2007}. In the present case very flat fcc Co platelets may be 
obtained under poor vacuum conditions\bracketsubfigref{fig-configs-special}a.

Let us compare the above figures for surface energy with data from the literature. Some data were reported for hcp Co, while others 
pertain to fcc Co. Besides, experimental values are measured at elevated temperature while theoretical ones are often relevant for 
zero temperature. Thus, a quantitative comparison is difficult. If one were interested in absolute values, and assuming hcp (0001) 
and fcc (111) planes have similar surface energies,  $\gamma_{111}\approx\unit[2.77]{\joule\per\meter\squared}$ based on 
calculations\cite{bib-VIT1998}, while experiments suggest 
$\gamma_{111}\approx\unit[2.53]{\joule\per\meter\squared}$\cite{bib-MIE1981,bib-deBOE1988}. However here we will cautiously 
restrict the discussion to energies normalized with $\gamma_{111}$ as a reference. There exists no prior quantitative discussion of 
fcc Co surface energies based on supported nanocrystals. Silly \etal\cite{bib-SIL2005b} simply report the existence of both 
$\{111\}$ and $\{100\}$ facets for Co nanocrystals on $\mathrm{SrTiO}_3(001)$, and Benamara \etal report $\{111\}$, $\{100\}$ and 
$\{110\}$ facets for fcc $\mathrm{Co}(001)$ dots self-assembled on $\mathrm{MgO}(001)$\cite{bib-BEN2010}.

It is often argued that surface energy increases for orientations with more open planes, so that we would expect 
$\gamma_{111}<\gamma_{100}<\gamma_{110}$\cite{bib-ALD1992}. Ignoring all details in the anisotropy of the band structure, this is 
expected as long as non-magnetic materials are considered. However non-zero spin polarization modifies this picture and is for 
instance responsible for a dip in the magnitude of surface energy towards the center of the $3d$ series, \ie for the strongest 
value of local moment\cite{bib-ALD1994}. A fine point is the following: as magnetic moments at surfaces are modified with respect 
to the bulk, in a manner that depends on the local coordination\cite{bib-GAM2003} and thus of the openness of this surface, the 
reduction of magnitude depends on the surface considered\cite{bib-PUN2011}. It is now recognized theoretically that this causes an 
anomaly in magnetic materials, where the densest place of atoms is not necessarily associated with the lowest surface energy: 
$\gamma_{001}<\gamma_{110}$ for bcc Fe\cite{bib-VIT1998}, and $\gamma_{001}<\gamma_{111}$ for fcc Co\cite{bib-ALD1994}.

$\Delta\gamma=\gamma_\mathrm{S}-\gammaint$ is more difficult to discuss. First, $\gamma_\mathrm{S}$ is not a value known \apriori 
from bulk properties like $\gamma_i$, as it is the surface energy of a monolayer of pseudomorphic Co wetting the $\mathrm{W}(110)$ 
surface. Second, $\gammaint$ results not only from electronic contributions expected to be weak in the case of a Stranski-Krastanov 
growth mode, however also from strain energy related to the accommodation of lattice misfit between the two elements. Nevertheless, 
assuming $\gamma_\mathrm{S}$ to be similar to $\gamma_{111}$, provides a large value for the interfacial energy: 
$\gammaint\approx0.6\gamma_{111}$, probably of primarily elastic origin.

\section{Micromagnetic configurations}

\begin{figure}
  \begin{center}
  \includegraphics[width=82.120mm]{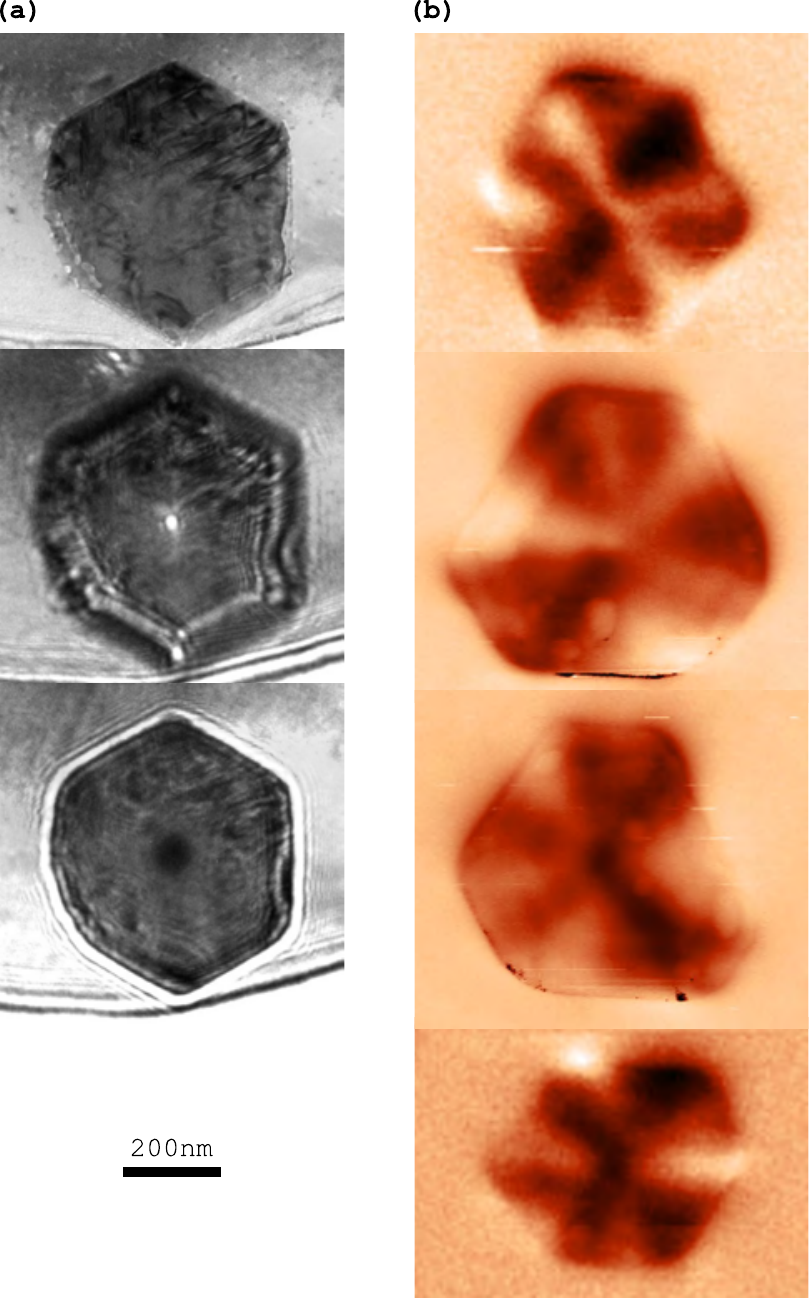}%
  \caption{\label{fig-configs-expe}(a)~Lorentz microscopy of an fcc Co dot. From top to bottom: in focus (structure only), 
over-focused and under-focused.\dataref{2008-09-12-Astro\"{\i}de et longueur paroi - Dot1} (b)~\dataref{AR66 - TP ENSPG2009}Magnetic 
Force Microscopy of several fcc Co dots. From top to bottom: vortex up and counter-clock wise (up-CCW), up-CW, down-CCW, down-CW.}
  \end{center}
\end{figure}

As fcc Co is a rather soft magnetic material, the critical single-domain size\cite{bib-HUB1998b,bib-FRU2005e} is a few times the 
dipolar exchange length, $\DipolarExchangeLength\approx\unit[5]{nm}$ for Co. The dots considered here have dimensions much above 
this, and are thus expected to display a flux-closure configuration. What type of pattern is formed is revealed with magnetic 
microscopy. We imaged the dots in the as-grown state with Lorentz and Magnetic Force Microscopy\bracketfigref{fig-configs-expe}. 
Whereas elongated dots of sufficient lateral size and thickness break into mainly two domains separated by a Bloch domain 
wall\cite{bib-FRU2003c}, here the contrast has a higher symmetry.

Lorentz microscopy unambiguously reveals the existence of a magnetic vortex at the center of the dot. The bright (or dark) contrast 
in over (or under) focal settings results from the opposite sign of the Lorentz force on either side of the vortex. This contrast 
informs us about the in-plane circulation of the flux closure, however not about the polarity of the vortex, \ie the sign of the 
vertical component of magnetization in its core. A strength of Lorentz microscopy is its high spatial resolution, which we used on 
such dots to investigate the dimensionality cross-over from a vortex to a Bloch wall\cite{bib-FRU2010c}.

\begin{figure}
  \begin{center}
  \includegraphics[width=85.555mm]{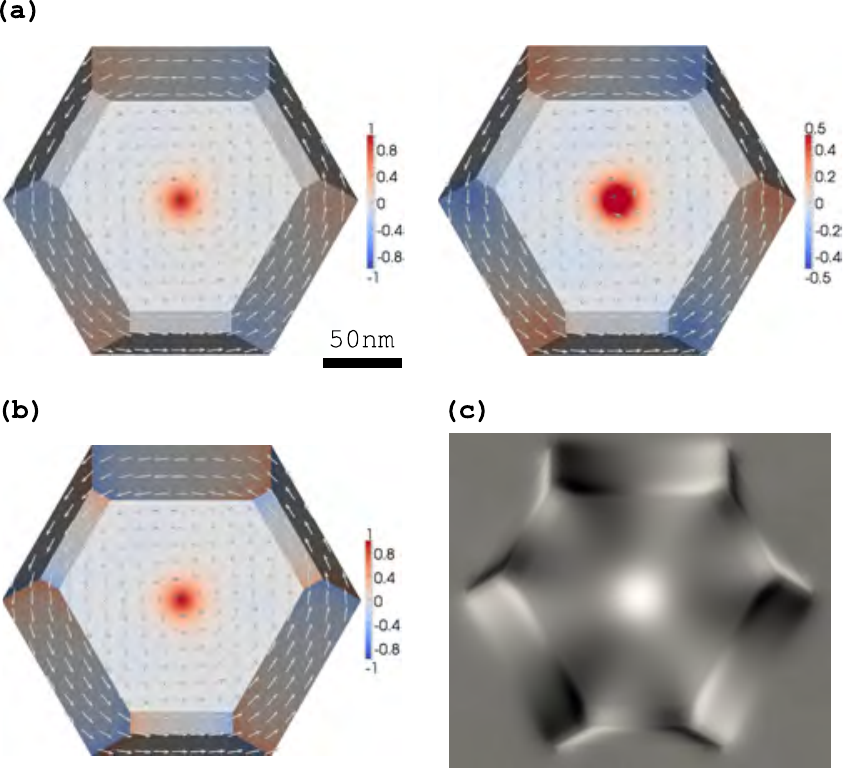}%
  \caption{\label{fig-configs-simul}\dataref{Simuls-Calculs-Astro\"{\i}de de chiralit\'{e}-FEM plot r\'{e}el (4-10-2010)}Micromagnetic 
simulations of an fcc Co(111) dot of thickness $\unit[50]{\nano\meter}$. (a)~Distribution of magnetization, with color coding the 
perpendicular component of surface magnetization, while the luminance revels the facets. Two ranges of colors are shown, suited to 
either the vortex, or the side facets. (b)~The color codes the surface charges (c)~Simulated MFM contrast: first order vertical 
derivative of the vertical component of stray field, calculated $\unit[10]{\nano\meter}$ above the surface of the dot.}
  \end{center}
\end{figure}

On the reverse, MFM is well suited to determine the polarity of the vortex. An up or down polarity is associated with opposite 
signs of the stray field above the dot, inducing a bright or dark monopolar contrast at the center of the 
dot\bracketsubfigref{fig-configs-expe}b. The contrast outside the vortex core is related to small-angle N\'{e}el walls, which give rise 
to a dipolar contrast. The sense of this contrast provides information about the in-plane circulation of magnetization around the 
core. The two degrees of freedom, core polarity and in-plane circulation, are independent\bracketsubfigref{fig-configs-expe}b. 
Those features are reproduced by micromagnetic simulations\bracketfigref{fig-configs-simul}. In addition, in the experiments the 
MFM contrast is shifted towards darker values, due to the so-called susceptibility contrast resulting from the softness of sample 
and/or tip\cite{bib-ABR1990}.

Let us notice a finer point. As magnetization remains mainly in-plane at the perimeter, the occurrence of surface charges cannot be 
avoided on the facets around the tilted edges\bracketsubfigref{fig-configs-simul}b. This gives rise to a sharp MFM contrast close 
to those edges. Due to the three-fold symmetry of the facets (inclination and difference of area on opposite sides), this 
contributes to the emergence of a three-fold symmetry of the contrast. While this is clear in the simulations, in the experiments 
it requires more attention due to the somewhat irregular shape. It is more marked for thicker dots as in 
\subfigref{fig-configs-expe}b where dipolar effects are more important.

\begin{figure}
  \begin{center}
  \includegraphics[width=86.836mm]{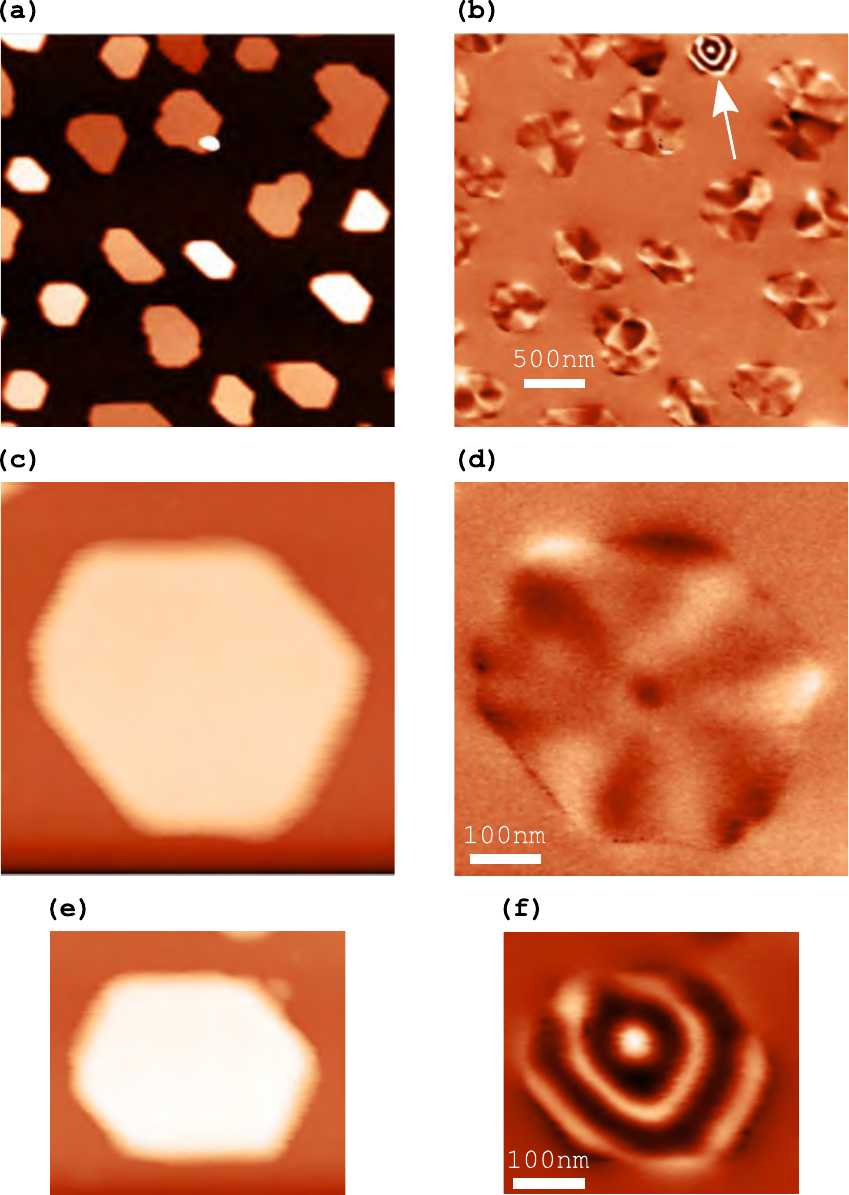}%
  \caption{\label{fig-configs-special}Co dots with a flat aspect ratio, and height in the range 
$[\unit[30\mathrm{-}60]{\nano\meter}]$. ~Large scale view of (a)~topography and (b)~MFM contrast. On the latter the white arrow 
highlights a dot with hcp lattice. (c-d)~Zoom of a fcc dot with height $\unit[33]{\nano\meter}$ (e-f)~Zoom of the hcp dot 
highlighted in (b), with height $\unit[25]{\nano\meter}$.}
  \end{center}
\end{figure}

We finally report special cases of self-assembled Co(111) dots. A first case is that of dots grown under poor vacuum conditions, 
which we already mentioned are much flatter than those reported above\bracketfigref{fig-configs-special}. For these the MFM 
contrast is sharper as the stray field is shorter-ranged. Although the signal over noise ratio is degraded, this decreases the 
contrast arising from the N\'{e}el walls and thus makes the inspection of the vortex core much easier than for thicker dots. A second 
case is dots with an hcp structure. Both Lorentz and MFM revealed a small fraction of dots with a drastically different 
distribution of magnetization with abundance of a few percent\bracketsubfigref{fig-configs-special}{a-b;e-f}, be they grown under 
high or poor vacuum conditions. This pattern is typical for stripe-domain phases, as found in materials with a bulk 
magneto-crystalline contribution to perpendicular magnetic anisotropy\cite{bib-KIT1946,bib-MAL1958,bib-MUR1966,bib-MUR1967}. In the 
case of finite-size systems the stripes tend to follow the edges and form a ring structure\cite{bib-HEH1996}. This view of local 
perpendicular magnetization is confirmed by the MFM contrast being much larger on such dots than on in-plane flux-closure ones. 
This suggests that these dots are hcp Co with the c axis perpendicular to the supporting surface. AFM confirms this, as the facets 
of such dots show a six-fold symmetry with an inclination of $\approx\angledeg{30}$, consistent with hcp$\{10-13\}$ facets.

To conclude, we used pulsed laser deposition to synthesize submicrometer Co fcc single crystals supported on a W(110)/Sapphire 
surface. From the analysis of the geometry of the dots we deduced the ratios of surface and interface energies of fcc~Co. $\{111\}$ 
surfaces have an energy more than ten percents higher than $\{100\}$ ones, resulting from the magnetic anomaly in surface energy. 
These dots form a flux-closure vortex state with an in-plane circulation of magnetization, suitable for studying vortex physics in 
a model system with no crystalline defects nor surface roughness.

\section*{Appendix~I}
\addcontentsline{toc}{section}{Appendix~I: angles for facets}%

\begin{figure}
  \begin{center}
  \includegraphics[width=84.344mm]{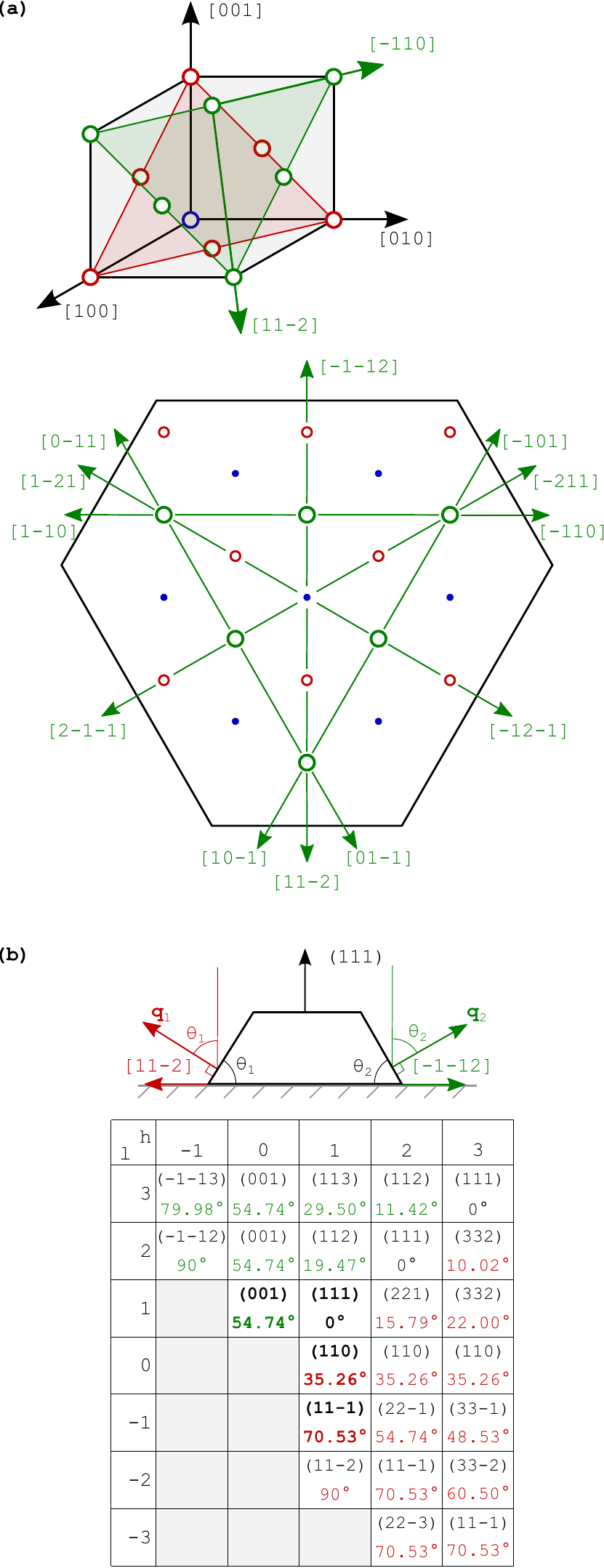}%
  \caption{\label{fig-facets-annex}(a)~Stacking of planes and directions of atomic rows for the fcc lattice. (b)~Notations for the 
identification of facets, and list of the main facets. Those observed experimentally are highlighted with bold figures.}
  \end{center}
\end{figure}

We calculate the inclination angles $\theta$ (with respect to the substrate plane) of the major possible facets of fcc(111) dots. 
We index facets with respect to their normal in the cubic reciprocal lattice: $\vect{q}=h\vect{a^*}+k\vect{b^*}+l\vect{c^*}$. As 
only angles matter here, for simplicity we release the fcc constraint that $h$, $k$ and $l$ should be of same parity. 
\figref{fig-facets-annex} shows the directions of atomic rows in the fcc lattice, and notations for the calculation. In the 
experiments facets only occur with <110> as the direction of the base, so that we restrict the calculation to this case.  In the 
following we consider $[-110]$ as the base atomic row, with no loss of generality. All facets obey $\vect q.[-110]=0$ and $\vect 
q.[111]\geq0$, implying $k=h$ and $l\geq-2h$. The facet angle then reads:

\begin{equation}
  \label{eq-angleFacets}\cos\theta=\frac{2h+l}{\sqrt{3}\sqrt{2h^2+l^2}}
\end{equation}

As facets on opposite sides are different due to the ABC stacking (see \subfigref{fig-facets-annex}a), we must distinguish the two 
cases: $\vect h_1.[11-2]\geq0$ and $\vect h_2.[-1-12]\geq0$, implying $l\geq h$ and $l\leq h$, respectively. The main facets are 
listed in \subfigref{fig-facets-annex}b.

\section*{Appendix~II}
\addcontentsline{toc}{section}{Appendix~II: Wulff construction}%

Here we provide relationships between the shape of a supported $\mathrm{fcc}(111)$ crystal and ratios of surface/interface 
energies. Using the notations illustrated in \figref{fig-facets} simple algebra yields geometrical ratios determined by 
surface/interface energies:

\begin{eqnarray}
  \label{eq-shapeVersusEnergy}
  \frac{h}{w} &=& \frac{1-\Delta\gamma}{\frac{3}{2\sqrt{2}}(1-\Delta\gamma)+\sqrt{\frac{3}{2}}\gamma_{100}}\\
  \frac{h_1}{h} &=& \frac{2\sqrt{\frac23}\gamma_{110}-1-\Delta\gamma}{1-\Delta\gamma}\\
  \frac{\ell_{111}}{\ell_{100}} &=& \frac{2\sqrt{2}\gamma_{100}-\sqrt{\frac32}(\Delta\gamma+1)}{\sqrt{2}(\sqrt{3}-\gamma_{100})}.
\end{eqnarray}

Alternatively and of interest in the present case, the determination of surface/interface energies from geometrical ratios of the 
crystal:

\begin{eqnarray}
  \label{eq-energyVersusShape}
  \gamma_{100} &=& \sqrt{3} \left[{ 1 - \frac{1}{ \left({1+\frac{\ell_{111}}{\ell_{100}}}\right) \left({1 - \frac hw 
\frac3{2\sqrt2}}\right) +1 } }\right]  \\
  \gamma_{110} &=& \sqrt{\frac32} \left[{ 1 - \frac3{2\sqrt2} \frac hw \frac{ \left({1+\frac{\ell_{111}}{\ell_{100}}}\right) 
\left({1-\frac{h_1}h}\right) }{ \left({1+\frac{\ell_{111}}{\ell_{100}}}\right) \left({1 - \frac hw \frac3{2\sqrt2}}\right) +1 
}}\right] \\
  \Delta\gamma &=& 1- \frac{3}{\sqrt2} \frac{h}{w} \frac{1 + \frac{\ell_{111}}{\ell_{100}}}{ 
\left({1+\frac{\ell_{111}}{\ell_{100}}}\right) \left({1 - \frac hw \frac3{2\sqrt2}}\right) +1 }.
\end{eqnarray}

\section*{Acknowledgments}

We acknowledge useful discussions with O.~Benamara (Cemes, Toulouse, France), V.~Dupuis~(LPMCN, Lyon, France), R.~Morel~(INAC, 
Grenoble, France) and technical help from A.~Rousseau, Z.~Kassir-Bodon, Ph.~David and V.~Guisset. This work was partly conducted as 
practicals for students, funded by Grenoble Institute of Technology~(G-INP). We acknowledge financial support from European FP6 
EU-NSF program (STRP 016447 MagDot).

\section*{References}


\end{document}